\documentclass[conference]{IEEEtran}
\usepackage{graphicx}
\usepackage{epstopdf}
\usepackage{cite}
\usepackage{amsmath,amssymb,amsfonts}
\usepackage{algorithmic}
\usepackage{graphicx}
\usepackage{textcomp}
\usepackage{xcolor}
\usepackage{comment}
\usepackage{amsthm} 
\usepackage{url}

\newtheorem{lemma}{Lemma}
\usepackage{comment}
\usepackage{amsthm}
\newtheorem{proposition}{Proposition}
\usepackage{flushend}

\def\BibTeX{{\rm B\kern-.05em{\sc i\kern-.025em b}\kern-.08em
    T\kern-.1667em\lower.7ex\hbox{E}\kern-.125emX}}
\begin{document}

\title{Version Age of Information with Contact Mobility in Gossip Networks\\

}

\author{Irtiza Hasan$\qquad$Ahmed Arafa\\Department of Electrical and Computer Engineering\\ University of North Carolina at Charlotte, NC 28223\\
\emph{ihasan@charlotte.edu}$\qquad$\emph{aarafa@charlotte.edu}
\thanks{This work was supported by the U.S. National Science Foundation under Grants CNS 21-14537 and ECCS 21-46099.}}

\maketitle

\begin{abstract}
A gossip network is considered in which a source node updates its status while other nodes in the network aim at keeping track of it as it varies over time. Information gets disseminated by the source sending status updates to the nodes, and the nodes gossiping with each other. In addition, the nodes in the network are mobile, and can move to other nodes to get information, which we term \emph{contact mobility.} The goal for the nodes is to remain as fresh as possible, i.e., to have the same information as the source's. To evaluate the freshness of information, we use the Version Age-of-Information (VAoI) metric, defined as the difference between the version of information available at a given node and that at the source. We analyze the effect of contact mobility on information dissemination in the gossip network using a Stochastic Hybrid System (SHS) framework for different topologies and mobility scalings with increasing number of nodes. It is shown that with the presence of contact mobility the freshness of the network improves in both ends of the network connectivity spectrum: disconnected and fully connected gossip networks. We mathematically analyze the average version age scalings and validate our theoretical results via simulations. Finally, we incorporate the cost of mobility for the network by formulating and solving an optimization problem that minimizes a weighted sum of version age and mobility cost. Our results show that contact mobility, with optimized mobility cost, improves the average version age in the network.
\end{abstract}

\section{Introduction}

Real-time status updating systems need to transmit time-sensitive information in order to make real-time inference. Information freshness is crucial for successful operations in autonomous vehicular systems\cite{Pappas2025}, wireless networks \cite{kadota2018}, Unmanned Aerial Vehichles (UAVs) and Internet of Things (IoT) networks\cite{choi2021} and many other emerging applications which can be modeled as gossip networks. 

Timeliness and information freshness are usually quantified using the Age-of-Information (AoI) metric \cite{kaul2012realtime}, which has found widespread usage in different systems \cite{yates2021age}. A variant of AoI, Version Age-of-Information (VAoI)\cite{yates2021gossip}, keeps track of the version of the source's information to quantify age. VAoI has been the main metric to quantify freshness in gossip networks using Stochastic Hybrid System (SHS) frameworks \cite{Mitra2022, CommunityGossip, Kaswan2023,Srivastava2025}, with other works studying it using methods other than SHS \cite{BFM, Bastopcu2022, Irtiza2025}. Our motivation to incorporate and analyze mobility in gossip networks stems from the fact that mobility is well known to improve performance in ad-hoc networks \cite{mobilitytse}, where the information transmission between source-destination (S-D) pairs using mobile intermediate relays is shown to enhance performance. It has been shown that mobility can improve freshness and timeliness in wireless networks in a similar S-D pair setting \cite{mobilityarafa}. In the gossip network we model and work with, there is a single source node and $n$ destination nodes. We expect mobility to significantly improve freshness in such network model as well.

\begin{figure}[t]
    \centering
    \includegraphics[width=1\linewidth]{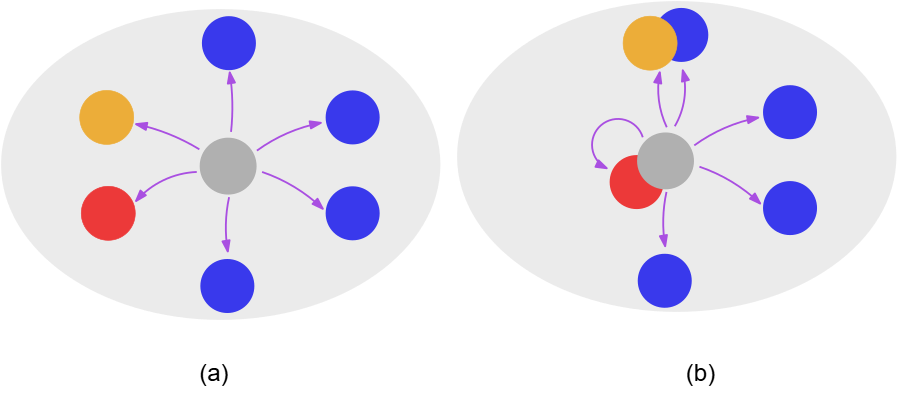} 
    \caption{(a) A disconnected network \textit{without} contact mobility. (b) A disconnected network \emph{with} contact mobility. Nodes can come in contact with other nodes in the network (orange) or the source node (red). The arrows indicate direct updates from the source to nodes; if the network is not disconnected, there would be update links between nodes as well.}
    \label{fig:mobility}
    \vspace{-.2in}
\end{figure} 

In fact, the effect of mobility in gossip networks has been studied in \cite{mobilityulukus} where the nodes in a gossip network exchange their positions pairwise according to a Poisson process. We term this \textit{exchange mobility} where ages of nodes switch based on position exchange. To show the effect of mobility scaling on the network, the mobility rate is parametrized as $\lambda_m$ = $\frac{\lambda}{f(n)}$ where $\lambda$ is the update rate from the source and $f(n)$ models how mobility rate decays with $n$. The study shows that exchange mobility lowers age and enhances its scaling properties in disconnected networks. However, in fully connected networks the results show that the system does not benefit from exchange mobility.

In this paper, we ask the following question: {\it instead of the nodes exchanging positions in the gossip network, what if nodes meet and exchange information, and update their VAoI accordignly?} We term this model of mobility \textit{contact mobility}, see Fig.~\ref{fig:mobility}. In contrast to the exchange mobility model, we no longer assume nodes exchange positions according to a Poisson process, rather we assume they meet each other according to one and retain the minimum age. We first derive equations of age in the network with contact mobility of nodes included and then we bound the age with different scalings of mobility (i.e., $f(n)$) with increasing number of nodes $n$. We also incorporate the cost of mobility into the framework and formulate an optimization problem to characterize the trade-off between age and cost mobility. Our results show that contact mobility \emph{always helps} in decreasing the age of nodes, even if the network is fully connected.

\section{System Model}

Consider a gossip network comprised of a source communicating with $n$ nodes. The source updates itself (receives a new information version) according to a Poisson process of rate $\lambda_e$. Let $\mathcal{N}=\{1,2,\dots,n\}$ denote the set of nodes in the network, and let the source node be denoted by $0$. The source pushes updates to node $j\in\mathcal{N}$ according to a Poisson process of rate $\lambda_{0j}$. In addition, nodes exchange information via two mechanisms. First, node $i\in\mathcal{N}$ can gossip with another node $j\in\mathcal{N}$, $j\neq i$, according to a Poisson process of rate $\lambda_{ij}$, provided they have a connection in the network topology. Second, nodes have {\it contact mobility}; a node $i\in\{0\}\cup\mathcal{N}$ can move to another node $j$, $j\neq i$, according to a Poisson process of rate $\lambda^m_{\{i,j\}}$, and upon meeting they share their information with each other. All four kinds of Poisson processes (source self-updates, source-to-nodes updates, nodes' gossiping and contact mobility) are assumed to be independent.

Let $N_s(t)$ and $N_i(t)$ denote the information version available at the source node and at node $i\in\mathcal{N}$, respectively. We define the Version Age-of-Information (VAoI), or merely age, at node $i$ as
\begin{align} \label{eq_vaoi}
  \Delta_i(t) = N_s(t) - N_i(t).
\end{align}
Thus, when node $i$ receives an update from the source directly at time $t$, its age resets to zero at $t^+$. Whereas when it receives a gossip update from node $j$ its age updates as the minimum of its own age and that of node $j$. On the other hand, when nodes $i$ and $j$ undergo contact mobility, {\it both} their ages are updated as
\begin{align}
    \Delta_i(t^+) = \Delta_j(t^+) = \min\{\Delta_i(t),\,\Delta_j(t)\}.
\end{align}

For any subset $S\subseteq\mathcal{N}$, let $N(S)$ denote the set of nodes that can send gossip updates into $S$, and $M(S)$ denote the set of nodes that can meet nodes in $S$ via contact mobility. In this work, we focus on the case in which the network has {\it full mobility}, where for every singleton $S=\{i\}$ we have
\begin{align}
    \bigl|\,M(\{i\})\cup\{i\}\bigr| = n,
\end{align}
where $|\cdot|$ denotes cardinality. This condition implies that any node in the network can come in contact with any other node.

\section{SHS Framework \& Main Results}

We utilize the Stochastic Hybrid Systems (SHS) framework \cite{hespanhashs} and follow the same framework and steps as in \cite{yates2021gossip} to derive the average version age $\tilde {v}_S$ of a subset $S$ of nodes in the network. 

The transitions in the system are described by a transition map $\mathcal{L}$ which is defined as
\begin{align}
\mathcal{L}
&= \{(0,0,\text{u})\} \nonumber\\
&\quad \cup\; \{(0,j,\text{u}):\, j\in\mathcal N\} \nonumber\\
&\quad \cup\; \{(i,j,\text{g}):\, i,j\in\mathcal N,\, i\neq j\} \nonumber\\
&\quad \cup\; \{(i,j,\text{m}):\, i,j\in \{0\}\cup\mathcal N,\, i\neq j\}.
\end{align}
From the transition map $\mathcal{L}$, there are four types of transitions.  First, the source can update itself, denoted as \((0,0,\text{u})\).  Second, the source can send an update to a node \(j\), represented by \((0,j,\text{u})\).  Third, gossiping between two nodes \(i\) and \(j\) is denoted by \((i,j,\text{g})\).  Finally, contact mobility between any two nodes \(i\) and \(j\) is represented by \((i,j,\text{m})\). The transition rates for each event type $(i,j,k)\in\mathcal{L}$ are given by
\begin{equation}
\lambda_{i,j,k} =
\begin{cases}
\lambda_e, 
& i=0,\ j=0,\ k=\text{u},\\[4pt]
\lambda_{0j}, 
& i=0,\ j\in\mathcal{N},\ k=\text{u},\\[4pt]
\lambda_{ij}, 
& i,j\in\mathcal{N},\ i\neq j,\ k=\text{g},\\[4pt]
\lambda^m_{\{i,j\}}, 
& \{i,j\}\subseteq \{0\}\cup\mathcal{N},\ i\neq j,\ k=\text{m}.
\end{cases}
\end{equation}
Here, $\lambda^m_{\{i,j\}}$ denotes the pairwise mobility rate between nodes $i$ and $j$. 

The state at time \(t\) is $\mathbf{X}(t)=[X_1(t),\dots,X_n(t)]$, where \(X_i(t)\) is the version age of node \(i\in\mathcal{N}\).  For a transition \((i,j,k)\in\mathcal{L}\), the post‐transition state \(\phi_{i,j,k}(\mathbf{X}(t))\) has the following values for its $l^{th}$ component:
\begin{equation}
X_l' =
\begin{cases}
X_l + 1, & \text{if } i=0,\ j=0,\ k=\text{u}, \\[6pt]
0, & \text{if } i=0,\ j=l,\ k = \text{u}, \\[6pt]
\min\{X_l,X_i\}, & \text{if } j=l,\ k = \text{g}, \\[6pt]
\min\{X_i,X_j\}, & \text{for } l \in \{i,j\},\ k = \text{m}, \\[6pt]
X_l, & \text{otherwise}.
\end{cases}
\end{equation}
We apply the extended generator of SHS to compute the dynamics.  Let 
\[
\psi_S(\mathbf{X}(t)) \;=\;\min_{i\in S}X_i(t).
\]
The extended generator is
\begin{equation}
(L\,\psi_S) (\mathbf{X}(t))
=\sum_{(i,j,k)\in\mathcal{L}}
\bigl[\psi_S(\phi_{i,j,k}(\mathbf{X}(t))) - \psi_S(\mathbf{X}(t))\bigr]\,
\lambda_{i,j,k}.
\end{equation}

By Dynkin’s formula, we have
\begin{equation}
\frac{d}{dt}\mathbb{E}[\psi_S(\mathbf{X}(t))]
=\mathbb{E}\bigl[L\,\psi_S(\mathbf{X}(t))\bigr].
\end{equation}

\begin{figure}[!t]
  \centering
  \includegraphics[width=0.5\textwidth]{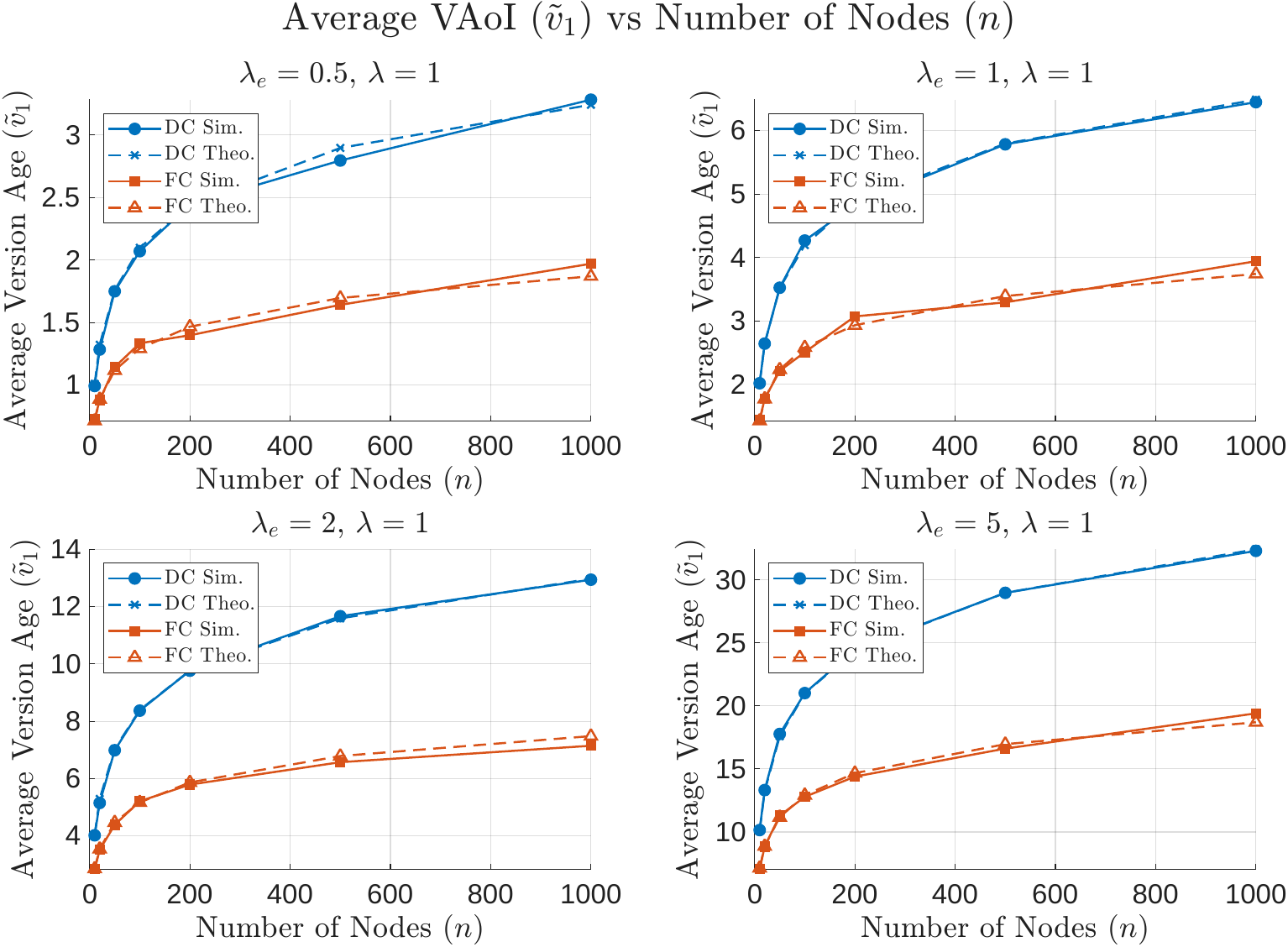}
  \caption{Theoretical and simulated values of average version age of a single node \(\tilde v_1\) with \(f(n)=n\) in DC and FC networks.}
  \label{fig:fnn}
\end{figure}

At steady state, the left‐hand side is zero. Substituting the expectation of the extended generator and letting $t \to \infty$, yields
\begin{align}
0 &= \lambda_e\bigl((\tilde v_S+1)-\tilde v_S\bigr)
    + \sum_{i\in S}\!\Big[\lambda_{0i}+\lambda^m_{\{0,i\}}\Big](0-\tilde v_S) \nonumber\\
  &\quad + \sum_{i\in N(S)}\;\sum_{j\in S}\lambda_{ij}\,(\tilde v_{S\cup\{i\}}-\tilde v_S) \nonumber\\
  &\quad + \sum_{i\in M(S)}\;\sum_{j\in S}\lambda^m_{\{i,j\}}\,(\tilde v_{S\cup\{i\}}-\tilde v_S).
\label{eq:eg}
\end{align}
Now let us define the total rates of the considered subset as
\begin{align}
\lambda_0(S)   &= \sum_{i\in S} \lambda_{0i}, 
&\quad \lambda^m_0(S) &= \sum_{i\in S} \lambda^m_{\{0,i\}}, \nonumber \\
\lambda_i(S)   &= \sum_{j\in S} \lambda_{ij}, 
&\quad \lambda^m_i(S) &= \sum_{j\in S} \lambda^m_{\{i,j\}}, \qquad i\notin S.
\end{align}

Rewriting \eqref{eq:eg} and solving for $\tilde v_S$ gives Proposition~\ref{thm_main}.

\begin{proposition} \label{thm_main}
In a mobile gossip network with contact mobility, for any subset of nodes $S \subseteq \mathcal{N}$, the steady‐state average version age $\tilde{v}_S$ is given by
\begin{equation}
\tilde{v}_S
=\frac{
\lambda_e 
+ \sum_{i\in N(S)}\lambda_i(S)\,\tilde v_{S\cup\{i\}}
+ \sum_{i\in M(S)}\lambda^m_i(S)\,\tilde v_{S\cup\{i\}}
}{
\lambda_0(S)
+ \sum_{i\in N(S)}\lambda_i(S)
+ \lambda^m_0(S)
+ \sum_{i\in M(S)}\lambda^m_i(S)
}.
\label{eq:proposition1}
\end{equation}
\end{proposition}

We note from the \eqref{eq:proposition1} that contact mobility between nodes can be interpreted as {\it additional gossip links} between them.

\begin{figure}[!th]
  \centering
  \includegraphics[width=1\linewidth]{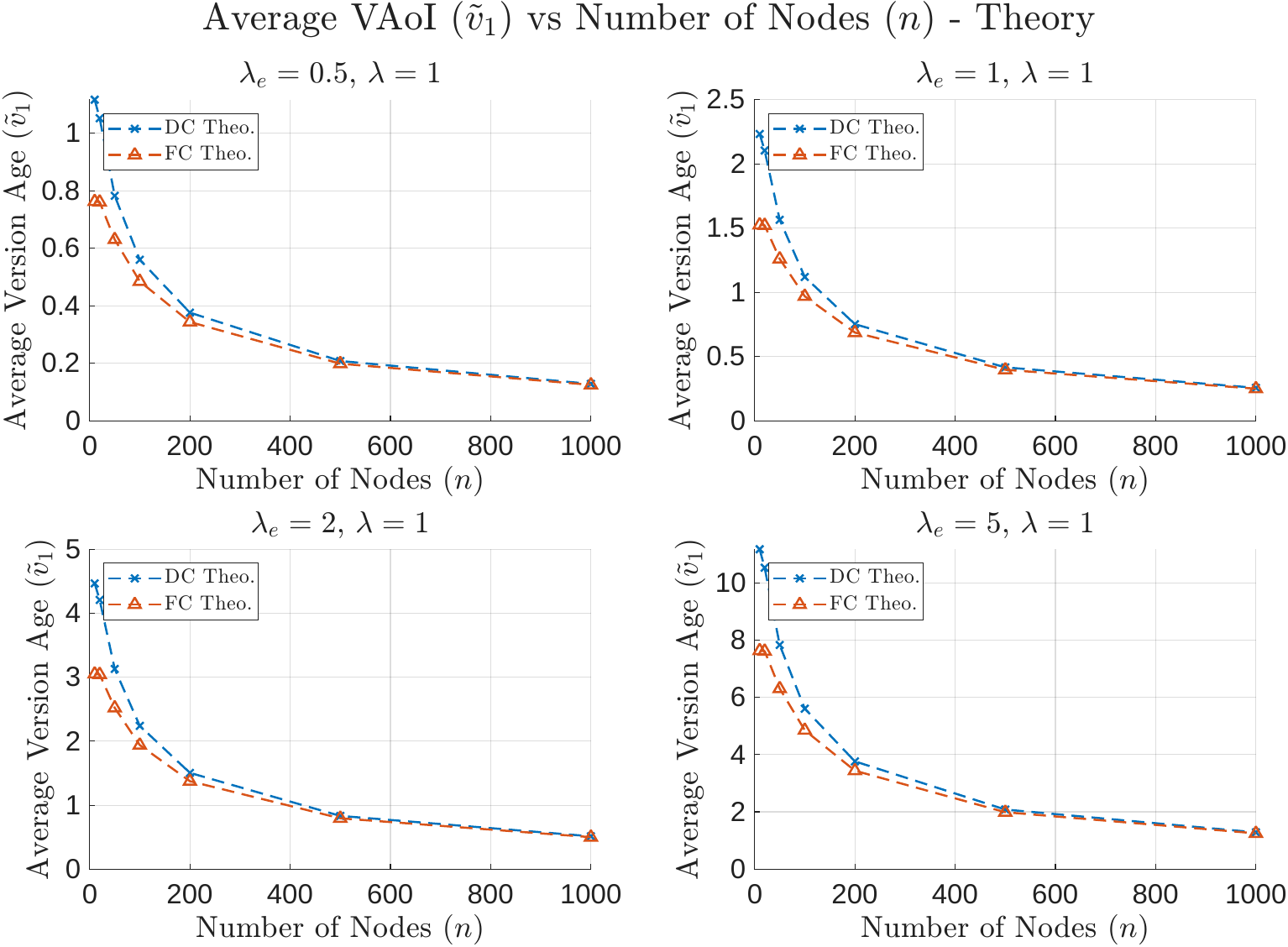}
  \caption{Theoretical \(\tilde v_1\) with \(f(n)=5\ln n\) in DC and FC networks.}
  \label{fig:fnln_t}
  \vspace{-.2in}
\end{figure}

\begin{figure}[!th]
  \centering
  \includegraphics[width=1\linewidth]{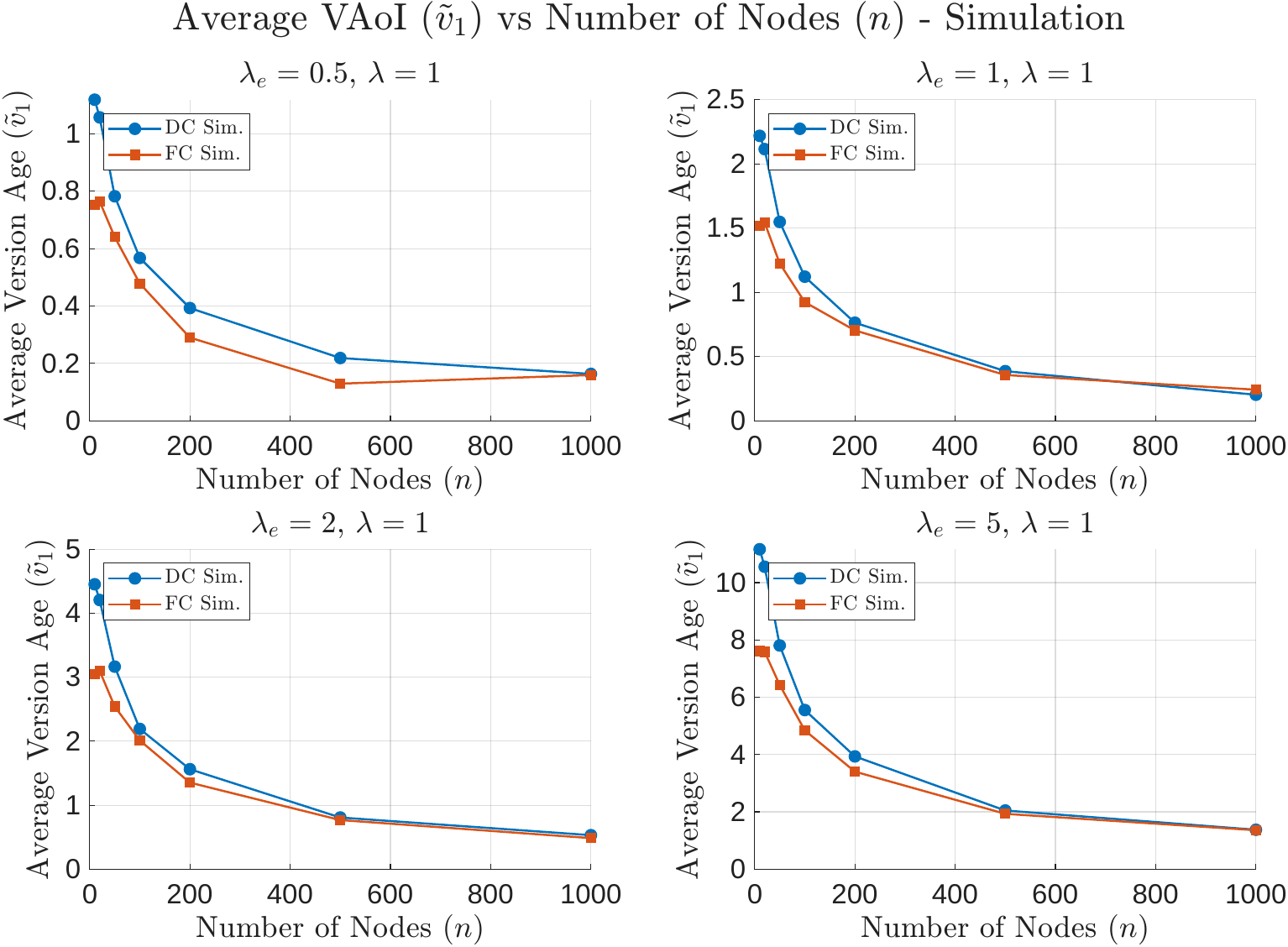}
  \caption{Simulated \(\tilde v_1\) with \(f(n)=5\ln n\) in DC and FC networks.}
  \label{fig:fnln_s}
  \vspace{-.1in}
\end{figure}

\begin{figure}[!t]
    \centering
    \includegraphics[width=1\linewidth]{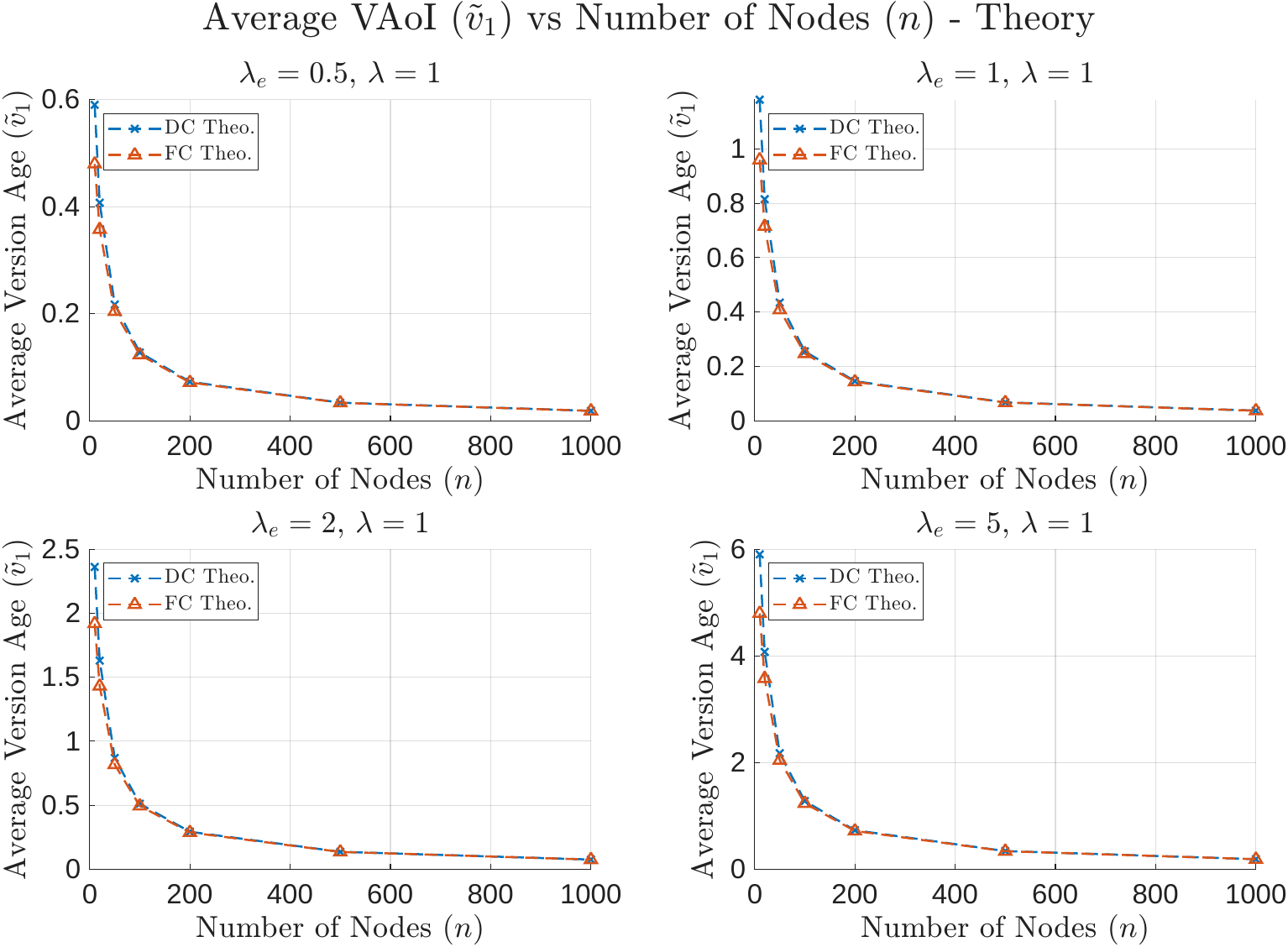} 
    \caption{Theoretical \(\tilde v_1\) with \(f(n)=5\) in DC and FC networks.}
    \label{fig:fnc_t}
    \vspace{-.2in}
\end{figure}

\begin{figure}[!t]
    \centering
    \includegraphics[width=1\linewidth]{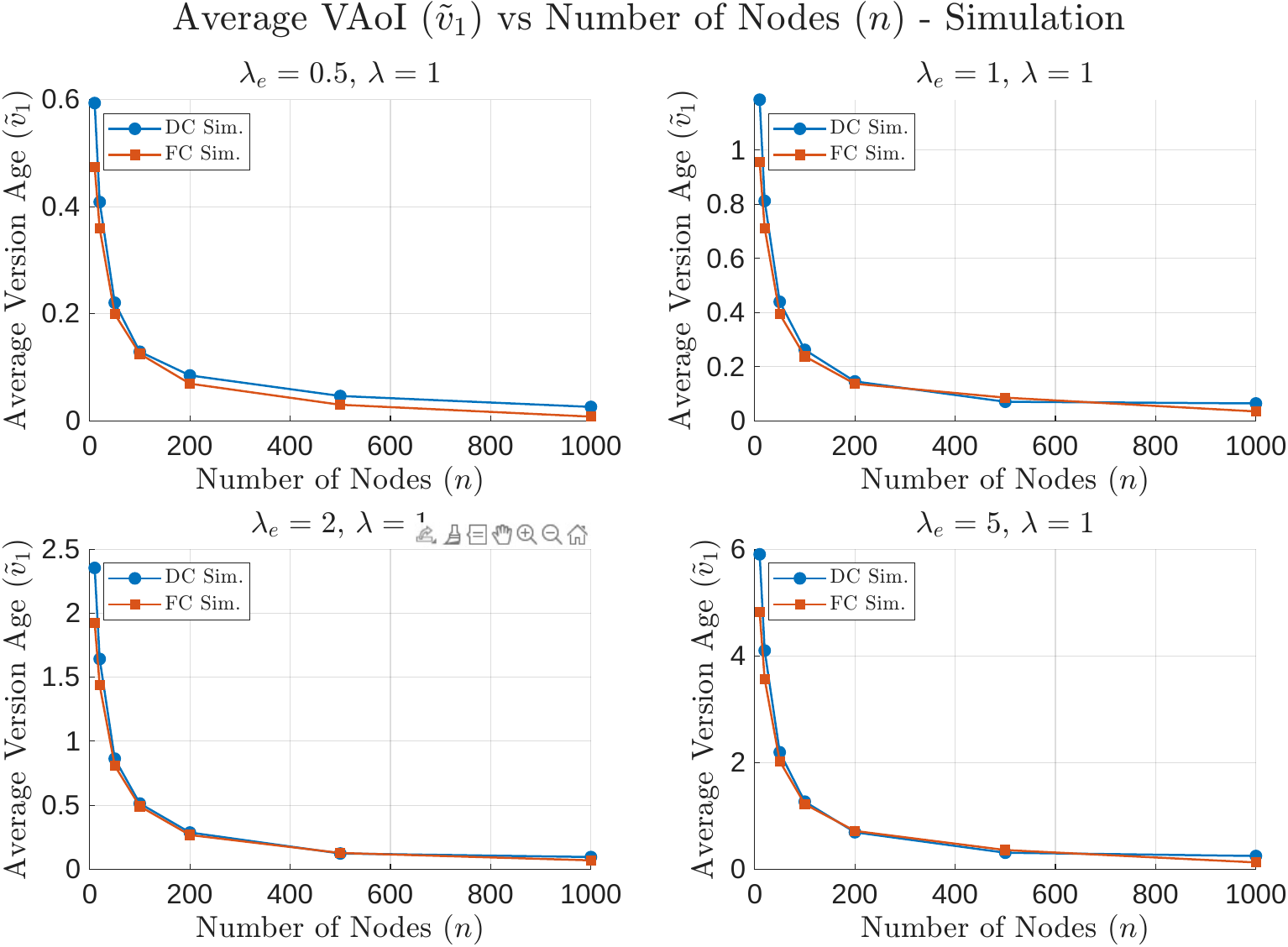} 
    \caption{Simulated \(\tilde v_1\) with \(f(n)=5\) in DC and FC networks.}
    \label{fig:fnc_s}
    \vspace{-.2in}
\end{figure}

From now on, we consider a symmetric setting in which the following holds: $\lambda_{0j}=\lambda/n,~\forall j \in \mathcal{N}$, and hence the overall source-to-nodes update rate is $\lambda$; $\lambda_{ij}=\lambda/(n-1),~\forall i,j \in \mathcal{N}$; and $\lambda_{\{i,j\}}^m=\lambda/f(n),~\forall i,j \in \{0\}\cup\mathcal{N}$. The function $f(n)$ controls the rate of mobility as the number of nodes grow. We focus on three main choices for $f(n)$: $n$, $c\ln n$, and $c$, for some $c>0$. We simulate gossip networks to validate the theoretical results employing an event-driven simulation to model the evolution of version age in both disconnected (DC) and fully connected (FC) network topologies. The results for simulation and corresponding theoretical values calculated using recursive methods are shown in Figs.~\ref{fig:fnn}, \ref{fig:fnln_t}, \ref{fig:fnln_s}, \ref{fig:fnc_t} and \ref{fig:fnc_s}. The parameter values used for the simulation are: $\lambda_e \in \{0.5, 1, 2, 5\}$ with $\lambda=1$ and $c=5$.

For \(f(n)=n\), the theoretical plots in Fig.~\ref{fig:fnn} are obtained by recursively solving \eqref{equationfig2dc} (cf. Appendix) for DC network and \eqref{equationfig2fc} for FC network in order to get $\tilde{v}_1$. We see that the average version age is significantly lower in FC network for different combinations of \(\lambda_e/\lambda\). Figs.~\ref{fig:fnln_t} and \ref{fig:fnln_s} consider \(f(n)=c\ln n\), where the theoretical curves are obtained from recursively solving \eqref{equationfig3dc} and \eqref{equationfig3fc} for the DC and FC networks, respectively. Simulated average age is consistent with the theoretical findings and one can notice the age of a FC network being lower than DC network. Finally, for $f(n)=c$, we use the recursions from solving \eqref{equationfig5dc} for the DC network and \eqref{equationfig5fc} for the FC network to get the results in Fig.~\ref{fig:fnc_t} which matches with simulated results in Fig.~\ref{fig:fnc_s}. In this case as well, the age in a FC network is slightly lower than DC network.

Next, we apply the above results to DC and FC network topologies in order to further analyze the impact of contact mobility on how version age scales with the number of nodes.

\section{Age Scaling with Contact Mobility}

In this section, we present an asymptotic analysis of how the version age scales with $n$ under contact mobility. We have the following lemma:

\begin{lemma} \label{thm-scaling}
In a gossip network with contact mobility and symmetric rates, for \emph{both} the DC and FC network topologies, the steady-state version age at any node scales as follows:
\begin{align}
    \tilde{v}_1 &= \mathcal{O}(\ln n), \quad \text{for $f(n)=n$}, \\
    \tilde{v}_1 &= \mathcal{O}\left(\frac{(\ln n)^2}{n}\right),\quad \text{for $f(n)=c\ln n,~c>0$}, \\
    \tilde{v}_1 &= \mathcal{O}\left(\frac{\ln n}{n}\right),\quad \text{for $f(n)=c,~c>0$}.
\end{align}
\end{lemma}

The proofs of all 6 cases contained in Lemma~\ref{thm-scaling}'s results are in the Appendix.

The results show that with contact mobility, the age scaling for both the DC and FC mobile networks is asymptotically the same. However, we argue that the FC network would still achieve a lower age because the additional gossip links provide more opportunities for information to spread.

Let us consider two parallel network systems operating with same sequence of random events for source updates and contact mobility meetings. In system A (mobile DC network), information spreads only through source transmissions and contact mobility. In system B (mobile FC network), information spreads through source transmissions, contact mobility, \textit{and} an additional independent set of Poisson processes representing the gossip links. At any point in time $t$, any node in system B has had every opportunity for an update that its counterpart in system A has had, \textit{plus} extra opportunities from the fixed gossip links. Because of these additional update opportunities, the version number of any node $i$ in system B, $N_i^B(t)$, must be greater than or equal to the version number of the corresponding node in system A, $N_i^A(t)$. Thus, $N_i^B(t) \ge N_i^A(t),~\forall i, t$. It follows by \eqref{eq_vaoi} that the age of any node in system B is always less than or equal to the age of its counterpart in system A:
\begin{align}
    \Delta_i^B(t) \le \Delta_i^A(t), ~\forall i, t.
\end{align}
This means that the steady-state average version age in the FC network will be lower than or equal to that in the DC network:
\begin{equation}
    \tilde {v}_{1}^{\text{FC}} \le \tilde {v}_{1}^{\text{DC}}.
\end{equation}

We note that similar arguments have been used  in \cite{Srivastava2025} where it is shown that the age in a push-pull network is lower that in push-only or pull-only networks due to more gossiping links. 

\vspace{-.1in}
\subsection{Contact Mobility vs.~Exchange Mobility}

\vspace{-.05in}

We observe that contact mobility provides a variant view point from the exchange mobility model introduced in \cite{mobilityulukus}. In exchange mobility, two nodes can exchange their positions according to a Poisson process of a certain mobility rate. Specifically, exchange mobility does not provide an improvement in age in symmetric networks. Take for example the DC network topology. Following the model in \cite{mobilityulukus}, one can show that for any node $j$, the steady-state average age is given by
\begin{align} \label{eq_exchange-mob-vj}
    \bar{v}_{j} = \frac{\lambda_e + \lambda_m(\bar{v}_1 + \bar{v}_2 + \dots + \bar{v}_{j-1} + \bar{v}_{j+1} + \dots + \bar{v}_n ) }{  \frac {\lambda}{n} + (n-1) \lambda_m  },
\end{align}
where $\lambda_m$ is the exchange mobility rate. The difference between the ages at the $j$th and the $i$th nodes can then be written as:
\vspace{-.1in}
\begin{equation}
\bar{v}_{j} - \bar{v}_{i} = \frac { \lambda_{m} \bar{v}_{i} }{  \frac {\lambda}{n} + (n-1) \lambda_m } - \frac { \lambda_{m} \bar{v}_{j} }{  \frac {\lambda}{n} + (n-1) \lambda_m },
\end{equation}
which further implies that
\begin{equation}
\bar{v}_{j} \left( 1 +  \frac { \lambda_{m} }{  \frac {\lambda}{n} + (n-1) \lambda_m }    \right) = \bar{v}_{i} \left( 1 +  \frac { \lambda_{m} }{  \frac {\lambda}{n} + (n-1) \lambda_m }    \right).
\end{equation}
Since $\left( 1 +  \frac { \lambda_{m} }{  \frac {\lambda}{n} + (n-1) \lambda_m }    \right) \neq 0$, we conclude that $ \bar{v}_j = \bar{v}_i,~\forall i,j,$ must hold in the DC network. Whence, \eqref{eq_exchange-mob-vj} can be written as
\begin{equation}
\bar{v}_j = \frac {\lambda_e + (n-1) \lambda_m \bar{v}_j }{ \frac {\lambda}{n} + (n-1)\lambda_m},
\end{equation}
which finally gives
\begin{equation}
\bar{v}_j = \frac {n \lambda_e}{\lambda}.
\end{equation}

Therefore, we get that $\bar{v}_j$ is {\it independent} of the mobility rate $\lambda_m$. In particular, even when $\lambda_m = 0$, the value of $\bar{v}_j$ remains $\frac{n \lambda_e}{\lambda}$. Thus, and as elaborated in \cite{mobilityulukus} for the FC network, exchange mobility does not reduce the version age in a DC network.

\begin{figure*}[!t]
    \centering
    \includegraphics[width=1\textwidth]{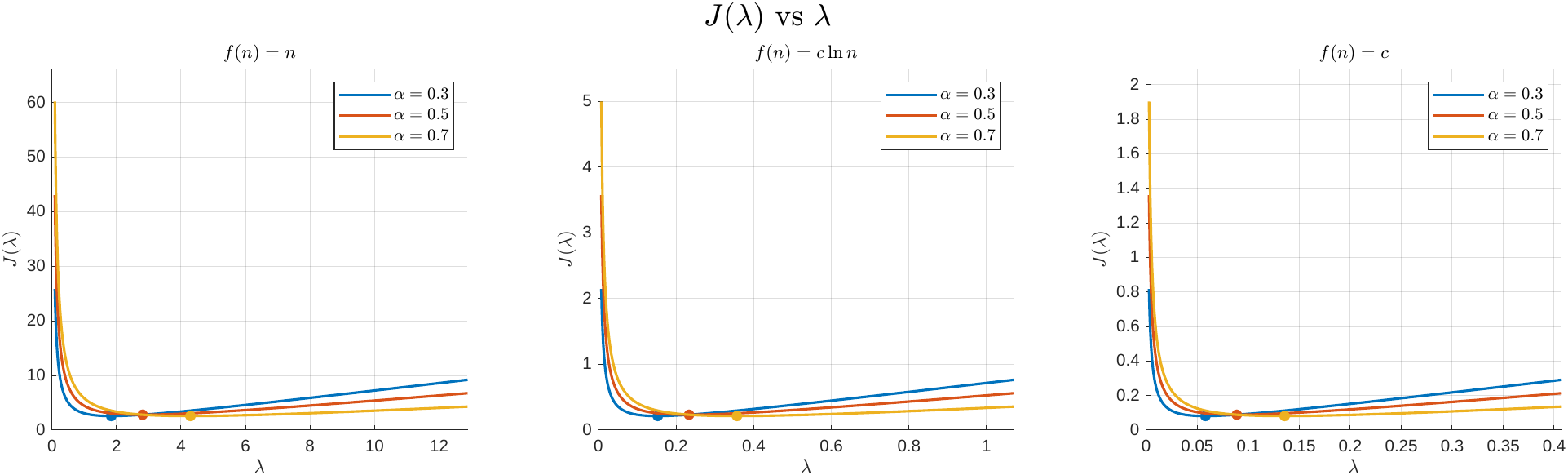}
    \caption{Trade-off curve between version age and mobility parameter for DC network. Each curve corresponds to a different value of $\alpha$; we set $n = 1000$, $\lambda_e = 1$ and $c = 1$.}
    \label{fig:mobilitycost}
\end{figure*}

We note that the DC network considered in \cite{mobilityulukus} is a 1-regular graph, meaning there is still some gossiping between adjacent nodes. For this kind of network, the age scales as
\begin{align}
    \mathcal{O}\bigl(\min\{f(n),\, \sqrt{n}\}\bigr).
\end{align}
That is, for $f(n)$ given by $n$, $ c \ln n$ and $c$, age scales as $\mathcal{O}(\sqrt{n})$, $\mathcal{O}(\ln n)$ and $\mathcal{O}(1)$, respectively. Since the network we consider in contact mobility modeling is strictly disconnected, we can conclude that the contact mobility DC network is working under more stringent conditions. Thus, one can compare the scalings above to those of Lemma~\ref{thm-scaling} for different values of $f(n)$ and conclude that contact mobility provides a better age improvement than exchange mobility. This, however, comes at the expense of a mobility cost, which we discuss next.

\section{Mobility Cost Optimization}

We define a joint cost function $J_{\alpha}(\lambda)$ to analyze the trade-off between the average version age $\tilde{v}_1(\lambda)$ and the rate parameter $\lambda$, which represents both mobility and update rates. We define the cost function as:
\begin{equation}
    J_{\alpha}(\lambda) = \alpha \tilde{v}_1(\lambda) + (1-\alpha)\lambda,
\end{equation}
for some $\alpha\in(0,1)$. The intuition is that lower values of $\alpha$ emphasize higher average mobility costs, and vice versa. Our goal is to find $\lambda^*$ that minimizes $J_{\alpha}(\lambda)$.

To get a handle on $J_{\alpha}$, we consider an upper bound on $\alpha \tilde{v}_1(\lambda)$ for the DC network. Such bound will also work for the FC network since it achieves a lower age. We assert that
\begin{equation}
    \tilde{v}_1(\lambda) \le \frac{K}{\lambda},
\end{equation}
for some specific values $K$ given by
\begin{align}
    K=\begin{cases}\lambda_e(\ln n + 1),&\text{if $f(n)=n$}\\
    \frac{c\lambda_e (\ln n)^2}{n}
              + \frac{c\lambda_e \ln n}{n}
              + \frac{c\lambda_e \ln n}{n(n + c \ln n)},&\text{if $f(n)=c \ln n$}\\
              \frac{c\lambda_e(\ln n + 1)}{n} + \frac{c\lambda_e}{n(c+n)},&\text{if $f(n)=c$}\end{cases},
\end{align}
where the above equations are direct implications of \eqref{eq:K1}, \eqref{eq:K2_1}, \eqref{eq:K2_2} and \eqref{eq:K3}  in Appendix.

With some abuse of notation, we now minimize the following cost function:
\begin{equation}
    J_\alpha(\lambda) = \alpha \frac{K}{\lambda} + (1-\alpha)\lambda. \label{eq:general_cost}
\end{equation}
Observe that $J_\alpha(\lambda)$ is convex in $\lambda$. Hence, setting the derivative to $0$ gives
\begin{equation}
   \lambda^* = \sqrt{\frac{\alpha K}{1-\alpha}}.
\end{equation}
Therefore, the minimum cost $J_\alpha(\lambda^*)$ is found by substituting $\lambda^*$ back into \eqref{eq:general_cost}:
\begin{align}
    J_\alpha(\lambda^*) 
                 &= 2\sqrt{\alpha(1-\alpha)K}. \label{eq:optimal_J_general}
\end{align}
For the specific expression for $K$, \eqref{eq:optimal_J_general} gives the optimal cost for a given choice of $\alpha$.

The trade-offs between mobility cost and age are visualized in Fig.~\ref{fig:mobilitycost}, which plots $J_\alpha(\lambda)$ versus $\lambda$ for different values of $\alpha$ across various network scenarios. We notice that in case of $f(n)= c$, the cost $J_\alpha$ is minimized for a very small value of $\lambda$. The optimal value of $\lambda$ increases as we move to the case of $f(n) = c\, \ln\, n\,$ and then to the case $f(n) = c\,$. The minimum cost achievable also has the same pattern where we get the lowest cost in the case of $f(n) = c$, followed by $f(n) = c\, \ln\, n\,$, and for $f(n) = n\,$ the cost at minima is highest.

\section{Conclusion}

We analyzed how contact mobility effects the Version Age of Information (VAoI) in gossip networks. By employing the Stochastic Hybrid Systems (SHS) framework, we derived analytical models for the average version age under various network topologies and mobility scaling scenarios. Our analysis shows that contact mobility always helps in lowering age, even for fully connected networks, which is in contrast to exchange mobility. Finally, we formulated and analyzed a cost function that balances the gains in information freshness against the cost of mobility and updates. By minimizing an upper bound on this cost, we derived the optimal rate that balances the trade-off between freshness and mobility.

\section*{Appendix}

In this appendix we provide detailed proofs for the 6 cases discussed in Lemma~\ref{thm-scaling}.

\subsection{Disconnected Network with $f(n) = n$}

\begin{proof}

We begin with the recursion in \eqref{eq:proposition1} for the average age $\tilde{v}_j$ of a subset of $j$ nodes. We have
\begin{equation}
\tilde{v}_j = \frac{\lambda_e + j(n-j)\frac{\lambda}{n} \tilde{v}_{j+1}}{j\frac{\lambda}{n} + j\frac{\lambda}{n} + j(n-j)\frac{\lambda}{n}}.
\label{equationfig2dc}
\end{equation}
Note that for a DC network we have $\lambda_i(S)=0,~\forall i$. Multiplying the numerator and denominator by $\frac{n}{j\lambda}$, we get
\begin{equation}
\tilde{v}_j = \frac{\frac{n\lambda_e}{j\lambda} + (n-j) \tilde{v}_{j+1}}{n-j+2}.
\label{eq:linear_recurrence}
\end{equation}
From this, $\tilde{v}_1$ can be found recursively as follows:
\begin{equation}
\tilde{v}_1 = \sum_{k=1}^{n-1} \left( \frac{\frac{n\lambda_e}{k\lambda}}{n-k+2} \right) \prod_{i=1}^{k-1} \frac{n-i}{n-i+2} + \tilde{v}_n \prod_{i=1}^{n-1} \frac{n-i}{n-i+2}.
\end{equation}
The product is evaluated as:
\begin{align}
\prod_{i=1}^{k-1} \frac{n-i}{n-i+2} =  \frac{(n-k+1)(n-k+2)}{n(n+1)}
\end{align}
Substituting this and noting that $\tilde{v}_n = \frac{\lambda_e}{2\lambda}$ (acquired directly from setting $j=n$ in \eqref{equationfig2dc}) gives:

\begin{equation}
\tilde{v}_1 = \frac{\lambda_e}{\lambda(n+1)} \sum_{k=1}^{n-1} \frac{n+1-k}{k} + \frac{\lambda_e}{\lambda n(n+1)}
\end{equation}
Next, we evaluate the summation:
\begin{equation}
\sum_{k=1}^{n-1} \frac{n+1-k}{k} = \sum_{k=1}^{n-1} \left(\frac{n+1}{k} - 1\right) = (n+1)H_{n-1} - (n-1),
\end{equation}
where $H_n$ denotes the $n$th harmonic number. Substituting this result back into the expression for $\tilde{v}_1$:
\begin{align}
\tilde{v}_1 &= \frac{\lambda_e}{\lambda(n+1)} \left( (n+1)H_{n-1} - (n-1) \right) + \frac{\lambda_e}{\lambda n(n+1)} \nonumber \\
&= \frac{\lambda_e}{\lambda}H_{n-1} - \frac{\lambda_e(n-1)}{\lambda(n+1)} + \frac{\lambda_e}{\lambda n(n+1)}.
\end{align}
Using the bound $H_{n-1} \le \ln n + 1$, and noticing the other terms result in a negative value for $n \geq$ 2, we have
\begin{equation}
    \tilde{v}_1 \leq \frac {\lambda_e}{\lambda}(\ln n + 1)
    \label{eq:K1}
\end{equation} 
We conclude that the average age of a node scales as
\begin{equation}
    \tilde{v}_1 = \mathcal{O}(\ln n).
\end{equation}
\end{proof}

\subsection{Fully Connected Network with $f(n) = n $}

\begin{proof}
We begin from \eqref{eq:proposition1}, for the average age $\tilde{v}_j$ of a subset of $j$ nodes in the FC network. We have
\begin{equation}
\tilde{v}_{j}=\frac{\lambda_{e}+j(n-j)\frac{\lambda}{n-1}\tilde{v}_{j+1}+j(n-j)\frac{\lambda}{n}\tilde{v}_{j+1}}{\frac{j\lambda}{n}+\frac{j(n-j)\lambda}{n-1}+\frac{j\lambda}{n}+\frac{j(n-j)\lambda}{n}}.
\label{equationfig2fc}
\end{equation}
Defining $\lambda_{\text{eff}} = \frac{\lambda}{n-1} + \frac{\lambda}{n}$, we get
\begin{equation}
\tilde{v}_{j}=\frac{\lambda_{e}+j(n-j)\lambda_{\text{eff}}\tilde{v}_{j+1}}{\frac{2j\lambda}{n}+j(n-j)\lambda_{\text{eff}}}
\end{equation}
Dropping the positive term $\frac{2j\lambda}{n}$ from the denominator, we get an upper bound:
\begin{align}
\tilde{v}_{j} &\le \frac{\lambda_{e}+j(n-j)\lambda_{\text{eff}}\tilde{v}_{j+1}}{j(n-j)\lambda_{\text{eff}}} \nonumber \\
&= \frac{\lambda_{e}}{j(n-j)\lambda_{\text{eff}}} + \tilde{v}_{j+1}.
\end{align}
Repeating this recursively, one can get that
\begin{align}
\tilde{v}_1 \le \frac{\lambda_e}{\lambda_{\text{eff}}} \sum_{k=1}^{n-1} \frac{1}{k(n-k)} + \tilde{v}_n 
\end{align}
Observe that the sum term can be simplified as follows:
\begin{align}
\sum_{k=1}^{n-1} \frac{1}{k(n-k)} &= \frac{1}{n} \sum_{k=1}^{n-1} \left(\frac{1}{k} + \frac{1}{n-k}\right) \nonumber \\
&= \frac{2H_{n-1}}{n}.
\end{align}
Upon noting that $\tilde{v}_{n}= \frac{\lambda_e}{2\lambda}$ (acquired directly from setting $j=n$ in \eqref{equationfig2fc}), we get that
\begin{equation}
    \tilde{v}_1 \le   \frac{\lambda_e}{\lambda_{\text{eff}}} \frac{2H_{n-1}}{n} + \frac{\lambda_e}{2\lambda}
\end{equation}

Next, substituting $\lambda_{\text{eff}} = \lambda \frac{2n-1}{n(n-1)}$ we have
\begin{align}
\tilde{v}_1 &\le \frac{\lambda_e}{\lambda \frac{2n-1}{n(n-1)}} \frac{2H_{n-1}}{n} + \frac{\lambda_e}{2\lambda} \nonumber \\
&=  \frac{\lambda_e}{\lambda} \frac{2(n-1)H_{n-1}}{2n-1} + \frac{\lambda_e}{2\lambda} 
\end{align}
This shows that the average version age scales as
\begin{equation}
    \tilde{v}_1 = \mathcal{O}(\ln n).
\end{equation}

\end{proof}

\subsection{Disconnected Network with  $f(n) = c\, \ln\, n$}

\begin{proof}
  
Starting with the recursion in \eqref{eq:proposition1} for the DC network with $f(n) = c \ln n$, we have
\begin{equation}
\tilde v_j = \frac{ \lambda_e + \frac{j(n-j)\,\lambda}{c\,\ln n}\,\tilde v_{j+1} }{ \frac{j\lambda}{n} + \frac{j\lambda}{c\,\ln n} + \frac{j(n-j)\,\lambda}{c\,\ln n} }.
\label{equationfig3dc}
\end{equation}
Note that for a DC network we have $\lambda_i(S)=0,~\forall i$. Dropping the positive term $\frac{j\lambda}{n}$ from the denominator gives
\begin{equation}
\tilde v_j < = \frac{ \lambda_e + \frac{j(n-j)\,\lambda}{c\,\ln n}\,\tilde v_{j+1} }{ \frac{j(n-j+1)\lambda}{c \ln n} }.
\end{equation}
This can be expressed as
\begin{align}
\tilde v_j &< \frac{c\,\lambda_e\,\ln n}{j(n-j+1)\lambda} + \frac{n-j}{n-j+1}\,\tilde v_{j+1}.
\end{align}

Repeating the above recursively one gets that
\begin{align}
\tilde v_1 < &\sum_{k=1}^{n-1} \left[ \frac{c\,\lambda_e\,\ln n}{k(n-k+1)\lambda} \prod_{i=1}^{k-1}\frac{n-i}{n-i+1} \right] \nonumber\\
&+ \left(\prod_{i=1}^{n-1}\frac{n-i}{n-i+1}\right)\,\tilde v_n.
\label{eq:unrolled_direct}
\end{align} 
The value of $\tilde{v}_n$ is found by setting $j=n$ in \eqref{equationfig3dc}:
\begin{equation}
\tilde v_n = \frac{\lambda_e}{\frac{n\lambda}{n} + \frac{n\lambda}{c \ln n}} = \frac{c\,\lambda_e\,\ln n}{\lambda\,(n + c\,\ln n)}.
\end{equation}
Simplifying the product $\prod_{i=1}^{k-1}\frac{n-i}{n-i+1} = \frac{n-k+1}{n}$, we can write
\begin{equation}
\tilde{v}_1 < \frac{c\,\lambda_e\,\ln n \, H_{n-1}}{n\lambda} + \frac{1}{n} \left( \frac{c\,\lambda_e\,\ln n}{\lambda\,(n + c\,\ln n)} \right).
\label{eq:K2_1}
\end{equation}
Using the bound $H_{n-1} \le \ln n + 1$, the first term above is bounded as follows:
\begin{equation}
\frac{c\,\lambda_e\,\ln n H_{n-1}}{n\lambda} = \frac{c\,\lambda_e\,(\ln n)^2}{\lambda\,n} + \frac{c\,\lambda_e\,\ln n}{\lambda\,n}.
\label{eq:K2_2}
\end{equation}

Combining all these results gives:
\begin{equation}
\tilde{v}_1 \le \left( \frac{c\,\lambda_e\,(\ln n)^2}{\lambda\,n} + \frac{c\,\lambda_e\,\ln n}{\lambda\,n} \right) + \frac{c\,\lambda_e\,\ln n}{n\lambda\,(n + c\,\ln n)}.
\label{eq:K2}
\end{equation}
Thus, we conclude that the average age of a node scales as
\begin{equation}
    \tilde v_1 = \mathcal{O}\left(\frac{(\ln n)^2}{n}\right).
\end{equation}

\end{proof}

\subsection{Fully Connected Network with $f(n) = c\, \ln\, n$}

\begin{proof}
Starting with the recursion in \eqref{eq:proposition1} for the FC network with $f(n) = c \ln n$, we have 
\begin{equation}
\tilde v_j = \frac{ \lambda_e + \left( \frac{j(n-j)\,\lambda}{\,n-1\,} + \frac{j(n-j)\,\lambda}{\,c \ln n\,} \right) \tilde v_{j+1} }{ \frac{j\lambda}{n} + \frac{j\lambda}{c \ln n} + \left( \frac{j(n-j)\,\lambda}{\,n-1\,} + \frac{j(n-j)\,\lambda}{\,c \ln n\,} \right) }.
\label{equationfig3fc}
\end{equation}
We drop the terms $\frac{j\lambda}{n}$ and $\frac{j\lambda}{c \ln n}$ to get
\begin{equation}
\tilde v_j < \frac{ \lambda_e + \left( \frac{j(n-j)\,\lambda}{\,n-1\,} + \frac{j(n-j)\,\lambda}{\,c \ln n\,} \right) \tilde v_{j+1} }{ \frac{j(n-j)\,\lambda}{\,n-1\,} + \frac{j(n-j)\,\lambda}{\,c \ln n\,} }
\end{equation}
Defining $\lambda_{\text{eff}} = \frac{\lambda}{n-1} + \frac{\lambda}{c \ln n}$, the inequality simplifies to
\begin{align}
\tilde v_j &< \frac{\lambda_e}{j(n-j)\lambda_{\text{eff}}} + \tilde v_{j+1}.
\end{align}

Repeating the above recursively we get
\begin{equation}
\tilde{v}_1 < \sum_{k=1}^{n-1} \frac{\lambda_e}{k(n-k)\,\lambda_{\mathrm{eff}}} + \tilde{v}_n
\end{equation}
Using the harmonic series identity $\sum_{k=1}^{n-1} \frac{1}{k(n-k)} = \frac{2H_{n-1}}{n}$, the bound becomes
\begin{equation}
\tilde{v}_1 < \frac{2\lambda_e H_{n-1}}{n\lambda_{\mathrm{eff}}} + \tilde{v}_n.
\label{eq:main_bound}
\end{equation}

Now since
\begin{equation}
\lambda_{\text{eff}} = \lambda \left(\frac{1}{n-1} + \frac{1}{c \ln n}\right) = \lambda \left(\frac{c \ln n + n - 1}{c(n-1)\ln n}\right),
\end{equation}
we have
\begin{equation}
\frac{2\lambda_e H_{n-1}}{n\lambda_{\mathrm{eff}}} = \frac{2\lambda_e H_{n-1}}{n} \left( \frac{c(n-1)\ln n}{\lambda(c \ln n + n - 1)} \right).
\end{equation}
Using the bounds: $H_{n-1} \le \ln n + 1$, $n-1 < n$, and $c\, \ln\, n + n - 1 > n-1$ for $n \ge 2$, the following holds:
\begin{align}
\frac{2\lambda_e H_{n-1} \cdot c(n-1)\ln n}{n\lambda(c \ln n + n - 1)} &< \frac{2\lambda_e (\ln n + 1) \cdot c \cdot n \cdot \ln n}{n \lambda (n-1)} \\
&< \frac{2c\lambda_e (\ln n + 1)\ln n}{\lambda(n-1)}.
\end{align}

Finally, we get $\tilde{v}_n$ by setting $j=n$ in \eqref{equationfig3dc} and putting all this together to have
\begin{equation}
\tilde v_n = \frac{\lambda_e}{\frac{n\lambda}{n} + \frac{n\lambda}{c \ln n}} = \frac{\lambda_e}{\lambda + \frac{n\lambda}{c \ln n}} = \frac{c \lambda_e \ln n}{\lambda(c \ln n + n)}
\end{equation}
We conclude that
\begin{equation}
\tilde{v}_1 = \mathcal{O}\left(\frac{(\ln n)^2}{n}\right).
\end{equation}

\end{proof}

\subsection{Disconnected Network with $f(n) = c$}

\begin{proof}
    
As always, we start with the recursion in \eqref{eq:proposition1} for the average age $\tilde{v}_j$ of a subset of $j$ nodes for the case of a DC network with $f(n) = c$ to get
\begin{equation}
\tilde {v}_{j}=\frac{\lambda_{e}+\frac{j(n-j)\lambda}{c}\tilde {v}_{j+1}}{\frac{j\lambda}{n}+\frac{j\lambda}{c}+\frac{j(n-j)\lambda}{c}}.
\label{equationfig5dc}
\end{equation}
Dropping the positive term $\frac{j\lambda}{n}$ from the denominator gives
\begin{align}
\tilde {v}_{j} &\le \frac{\lambda_{e}+j(n-j)\frac{\lambda}{c}\tilde {v}_{j+1}}{j(n-j+1)\frac{\lambda}{c}} \\
&= \frac{c\lambda_{e}}{j(n-j+1)\lambda} + \frac{n-j}{n-j+1}\tilde{v}_{j+1}.
\end{align}
Repeating this recursively gives
\begin{equation}
\begin{split}
    \tilde{v}_{1} 
    \le\; &\sum_{k=1}^{n-1} \left( \frac{c\lambda_{e}}{k(n-k+1)\lambda} 
    \prod_{i=1}^{k-1} \frac{n-i}{n-i+1} \right) \\
    &+ \left( \prod_{i=1}^{n-1} \frac{n-i}{n-i+1} \right) \tilde{v}_{n},
\end{split}
\label{eq:recursive-bound}
\end{equation}
and upon simplifying the product terms we further have
\begin{align}
\tilde{v}_{1} &\le \sum_{k=1}^{n-1} \left( \frac{c\lambda_{e}}{k(n-k+1)\lambda} \cdot \frac{n-k+1}{n} \right) + \frac{1}{n}\tilde{v}_{n} \\
&\le \frac{c\lambda_{e}}{n\lambda} \sum_{k=1}^{n-1} \frac{1}{k} + \frac{\tilde{v}_{n}}{n}
\end{align}

We find $\tilde {v}_{n}$ by setting $j=n$ in \eqref{equationfig5dc} to get
\begin{equation}
    \tilde{v}_{n} = \frac{\lambda_{e}}{\lambda + \frac{n\lambda}{c}} = \frac{c\lambda_{e}}{\lambda(c+n)}.
\end{equation}
Substituting this back and noting that the summation represents the $(n-1)$th Harmonic number we have
\begin{align}
\tilde{v}_{1} &\le \frac{c\lambda_{e}}{n\lambda} H_{n-1} + \frac{1}{n} \left(\frac{c\lambda_{e}}{\lambda(c+n)}\right) \\
&\le \frac{c\lambda_{e}(\ln n + 1)}{n\lambda} + \frac{c\lambda_{e}}{n\lambda(c+n)}.
\label{eq:K3}
\end{align}
The age, therefore, scales as
\begin{align}
\tilde{v}_{1}=\mathcal{O}\left(\frac{\ln n}{n}\right).
\end{align}

\end{proof}

\subsection{Fully Connected Network with  $f(n) = c $}

\begin{proof}
The recursion in \eqref{eq:proposition1} for the case of a FC network with $f(n) = c$ gives
\begin{equation}
  \tilde v_j
  \frac{
      \displaystyle
      \lambda_e
      + \frac{j(n-j)\,\lambda}{n-1}\,\tilde v_{j+1}
      + \frac{j(n-j)\,\lambda}{c}\,\tilde v_{j+1}
    }{
      \displaystyle
      \frac{j\lambda}{n}
      + \frac{j(n-j)\,\lambda}{n-1}
      + \frac{j\lambda}{c}
      + \frac{j(n-j)\,\lambda}{c}
    }.
  \label{equationfig5fc}
\end{equation}
Defining $\lambda_{\mathrm{eff}}
  = \frac{\lambda}{n-1} \;+\; \frac{\lambda}{c}
$, we have
\begin{align}
  \tilde v_j
  &= \frac{
       \lambda_e
       + j(n-j)\,\lambda_{\mathrm{eff}}\,\tilde v_{j+1}
     }{
       \tfrac{j\lambda}{n}
       + \tfrac{j\lambda}{c}
       + j(n-j)\,\lambda_{\mathrm{eff}}
     }.
\label{eq:fc_c}
\end{align}
Dropping the term \(\tfrac{j\lambda}{n} + \tfrac{j\lambda}{c} \geq 0\), we get
\begin{align}
  \tilde v_j
   \leq& \frac{
         \lambda_e + j(n-j)\,\lambda_{\mathrm{eff}}\,\tilde v_{j+1}
       }{
         j(n-j)\,\lambda_{\mathrm{eff}}
       } \nonumber \\
   =& \frac{\lambda_e}{j(n-j)\,\lambda_{\mathrm{eff}}}
     + \tilde v_{j+1}.
  \label{eq:const-bound}
\end{align}

Applying this recursively, we have
\begin{align}
  \tilde v_1
  &\leq \sum_{k=1}^{n-1}
      \frac{\lambda_e}{k(n-k)\,\lambda_{\mathrm{eff}}}
    + \tilde{v}_n \nonumber \\
  &=\sum_{k=1}^{n-1}
      \frac{\lambda_e}{k(n-k)\,\lambda_{\mathrm{eff}}}
    + \frac{\lambda_e}
         {\lambda + n\bigl(\lambda/c\bigr)},
\end{align}
where we find $\tilde {v}_{n}$ by setting $j=n$ in \eqref{equationfig5fc}. Noting that
\(\displaystyle\sum_{k=1}^{n-1}\frac1{k(n-k)}=\tfrac{2H_{n-1}}{n}\), we get
\begin{align}
  \tilde v_1
  &\leq \frac{\lambda_e}{\lambda_{\mathrm{eff}}}
     \,\frac{2H_{n-1}}{n}
    + \frac{\lambda_e}
         {\lambda + n\bigl(\lambda/c\bigr)}\\[1ex]
  &\leq \frac{\lambda_e}{\lambda_{\mathrm{eff}}}
     \,\frac{2(\ln n + 1)}{n}
    + \frac{\lambda_e}
         {\lambda + n\bigl(\lambda/c\bigr)}.
    \qquad
\end{align}

Substituting back the value of $\lambda_{\mathrm{eff}}$, we get
\begin{align}
 \tilde v_1
 &\leq \frac{\lambda_e}{\lambda\left(\frac{1}{n-1} + \frac{1}{c}\right)} \cdot \frac{2(\ln n + 1)}{n}
    + \frac{\lambda_e}{\lambda\left(1 + \frac{n}{c}\right)} \\
 &= \frac{\lambda_e}{\lambda} \cdot \frac{c(n-1)}{c+n-1} \cdot \frac{2(\ln n + 1)}{n}
    + \frac{\lambda_e}{\lambda} \cdot \frac{c}{c+n} \\
 &= \frac{2c(n-1)\lambda_e(\ln n + 1)}{\lambda n (c+n-1)}
    + \frac{c\lambda_e}{\lambda(c+n)} \\
 &< \frac{2c\lambda_e(\ln n + 1)}{\lambda n}
    + \frac{c\lambda_e}{\lambda(c+n)}
\end{align}
It follows that age of a node $\tilde v_1$ scales as
\begin{align}
\tilde{v}_{1}=\mathcal{O}\left(\frac{\ln n}{n}\right).
\end{align}

\end{proof}

\end{document}